\documentclass[12pt]{article}
\usepackage{times,epsfig,amsmath}

\newcommand{\anc}{\rule{0mm}{0mm}}

\newcommand{\Eslash}{{\not{\!\!E}}}

\newcommand{\sq}{{\tilde{q}}}
\newcommand{\sqR}{{\tilde{q}_{\rm R}}}
\newcommand{\sqL}{{\tilde{q}_{\rm L}}}
\newcommand{\su}{{\tilde{u}}}

\newcommand{\suL}{{\tilde{u}_{\rm L}}}
\newcommand{\sd}{{\tilde{d}}}

\newcommand{\sdL}{{\tilde{d}_{\rm L}}}

\newcommand{\go}{\tilde{g}}

\newcommand{\cha}{\tilde{\chi}}
\newcommand{\neu}{\tilde{\chi}^0}

\begin{document}
\title{Testing the SUSY-QCD Yukawa coupling\\ in a combined LHC/ILC analysis}

\author{A. FREITAS$^{1}$ and P. Z. SKANDS$^{2}$}
\maketitle

\noindent
{\centering\it $^{1}$Institut f\"ur Theoretische Physik,
        Universit\"at Z\"urich, Winterthurerstrasse 190, CH-8057
        Z\"urich\\
$^{2}$Theoretical Physics, Fermi National Accelerator Laboratory, Batavia, IL
60510-0500, USA\\}

\begin{abstract}
In order to establish supersymmetry (SUSY) at future colliders, 
 the identity of gauge couplings
and the corresponding Yukawa couplings between gauginos, sfermions and
fermions needs to be verified. 
Here a first phenomenological study for determining 
the Yukawa coupling of the SUSY-QCD sector
is presented, using a
method which combines information from LHC and ILC.
\end{abstract}

\section{Introduction}

One of the fundamental relations in softly broken supersymmetric theories is the equality 
between the Yukawa
coupling $\hat{g}$ of a gaugino interacting with a fermion and a sfermion
and the corresponding Standard Model (SM) gauge coupling $g$ of a
gauge boson and two (s)fermions,
$
g = \hat{g}
$.
At colliders, this relation can be investigated through the  production cross-sections 
for SUSY particles. Within the Minimal Supersymmetric Standard Model (MSSM),
it has been shown \cite{susyid,ckmz,slepo,slep} that 
the SUSY Yukawa couplings in the electroweak
sector can be precisely tested at the per-cent level at a high-energy $e^+e^-$ collider (ILC).

However, the analysis of the SUSY Yukawa coupling $\hat{g}_{\rm s}$ 
in the QCD sector is much more difficult. 
At the ILC this interaction could be studied in
the process $e^+e^- \to q \sq^* \go$, $\bar{q} \sq \go$
\cite{BMWZ}, but suffers from very low rates and large backgrounds.
At the LHC on the other hand, squarks and gluinos with masses below
2--3 TeV are abundantly produced, and their pair production cross sections
depend directly on 
$\hat{g}_{\rm s}$. However, measurements of total cross sections are
exceedingly difficult at hadron colliders, with 
typically only one or two specific decay channels of the squarks and
gluinos experimentally accessible \cite{lhclc}. 

In this contribution, a combination of ILC and LHC measurements is considered, where the relevant branching ratios are to be
determined at ILC and combined with exclusive
cross section measurements in selected channels at the LHC.


\section{Squark production at the LHC}
\label{sc:lhc}

\begin{figure}
\centering
\anc\hspace{2mm}%
\psfig{figure=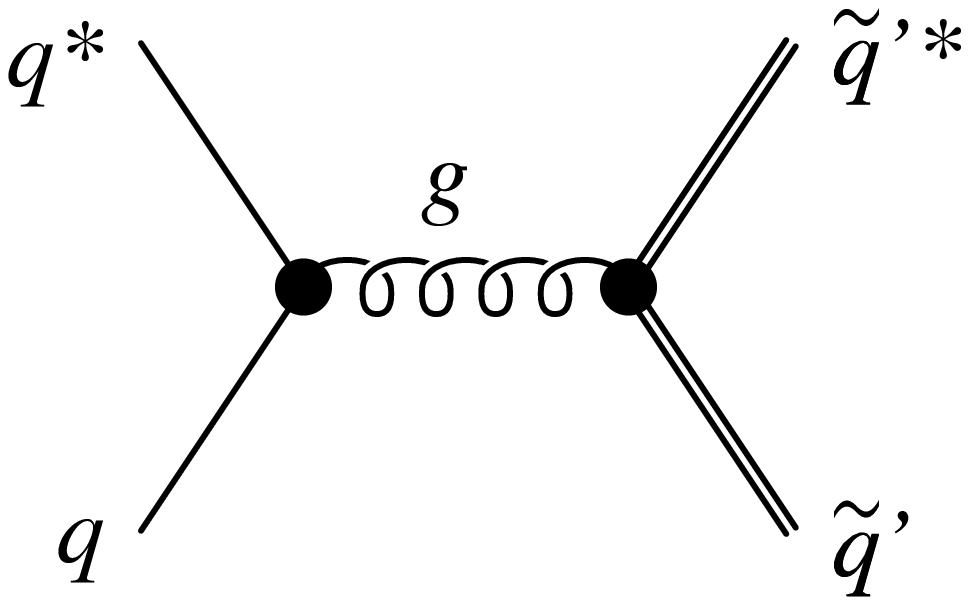, width=3cm}
\psfig{figure=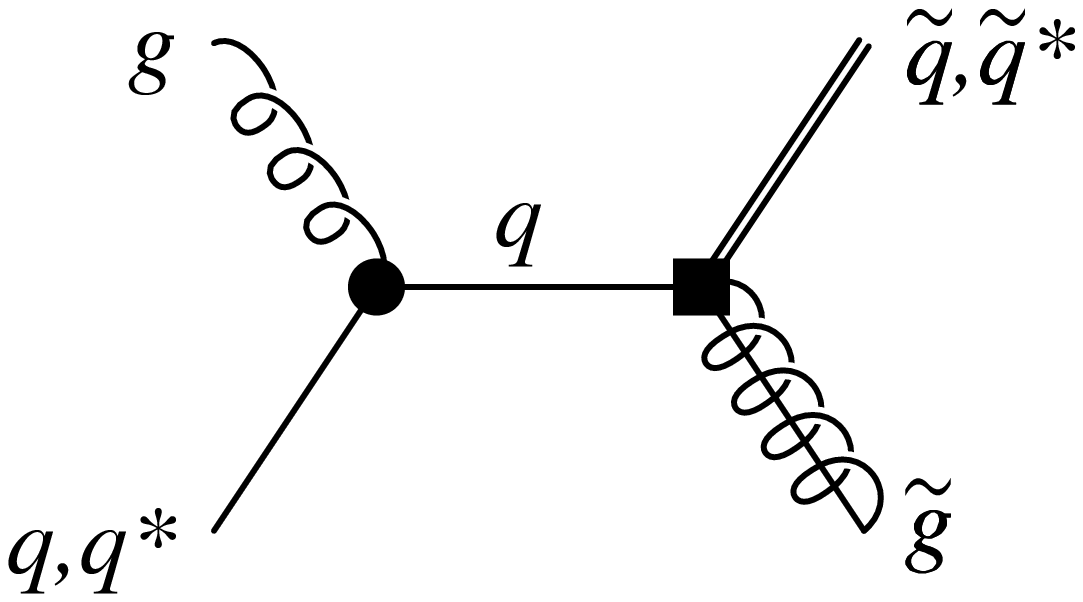, width=3.6cm}
\psfig{figure=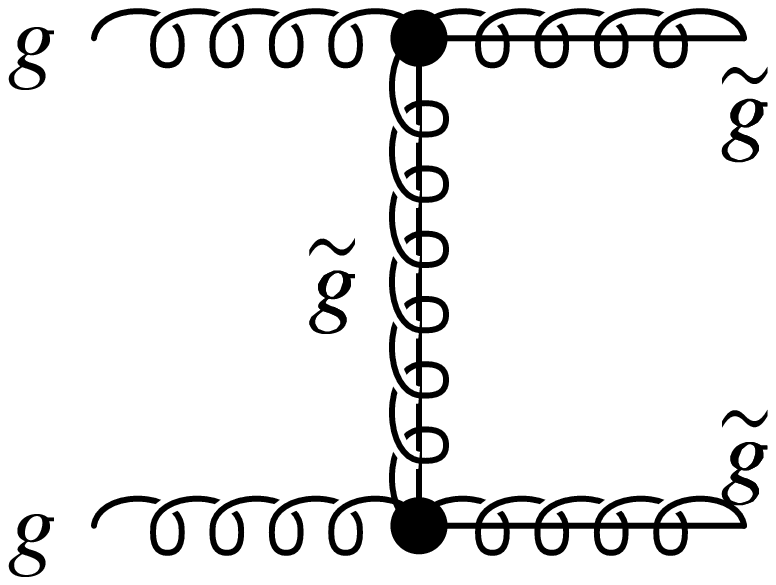, width=2.3cm}\\
\psfig{figure=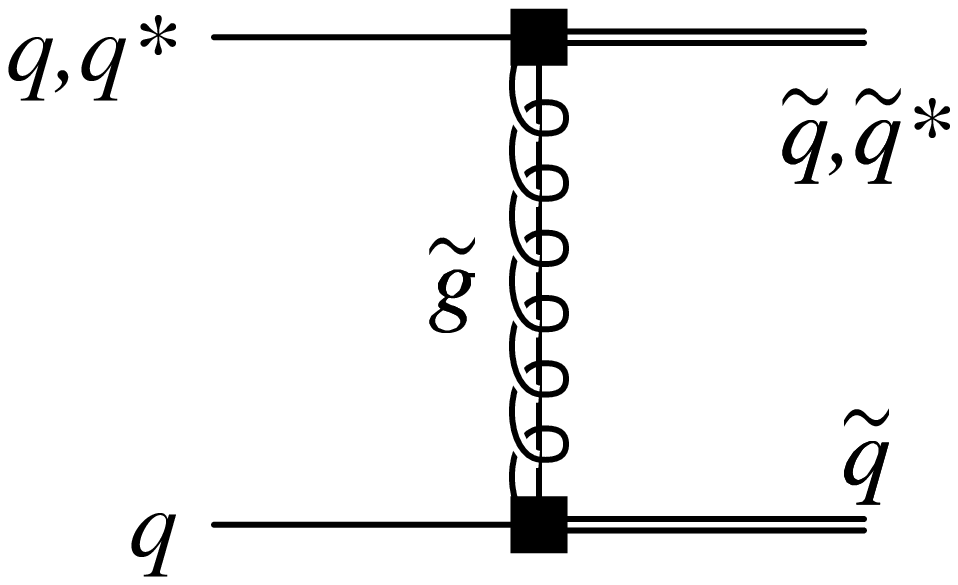, width=3.1cm}
\hspace{4mm}
\raisebox{1mm}{\psfig{figure=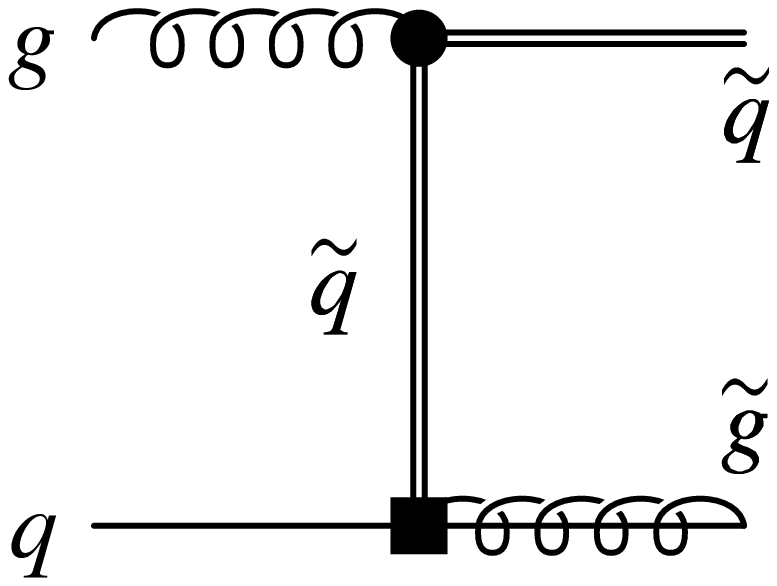, width=2.2cm}}
\hspace{7mm}
\raisebox{1mm}{\psfig{figure=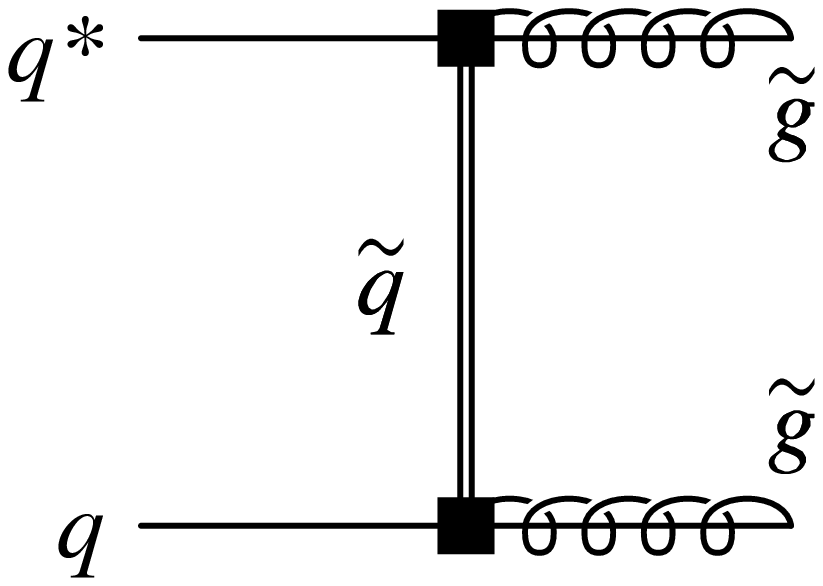, width=2.5cm}}
\caption{Examples for Feynman diagrams for 
partonic squark and gluino production in $pp$
collisions. Dots indicate the gauge coupling $g_{\rm s}$, squares 
the Yukawa coupling $\hat{g}_{\rm s}$.}
\label{fg:dia1}
\end{figure}

In $pp$ collisions, squarks and gluinos can be produced in various combinations, see Fig.~\ref{fg:dia1}.
The production of same-sign squarks 
($\suL\suL$, ...) 
is especially interesting, since it only
proceeds through the diagram in the lower left of
Fig.~\ref{fg:dia1}, 
and thus depends solely on the SUSY
Yukawa coupling $\hat{g}_{\rm s}$. In $pp$ collisions this process 
dominantly produces $\su$ and $\sd$ squarks, in direct proportion to the quark
content of the proton. 

While the typically lighter $\sqR$ almost exclusively decays directly into the lightest neutralino and a quark jet, thus not allowing charge tagging, for the heavier $\sqL$ 
the charge of the squark can be tagged through a chargino decay chain,
\begin{align}
\suL &\to d \, \cha^+_1 \to d \, l^+ \, \nu_l \, \neu_1, &
\sdL &\to u \, \cha^-_1 \to u \, l^- \, \bar{\nu}_l \, \neu_1
.
\end{align}
The production of same-sign squarks 
with this decay channel will thus lead to two
same-sign leptons, two hard jets and missing transverse energy in the
final state. Other direct squark
production processes on the other hand will tend to produce opposite-sign leptons. 

A very problematic background
can come from gluino pair and mixed gluino-squark production if $m_{\go} >
m_{\sqL}$. In this case, gluinos can decay into quarks and squarks, $\go
\to q \, \sqL$, generating a component of 
two same-sign squarks plus additional jets.  This
background is very challenging if the mass difference 
$m_{\go} - m_{\sqL}$ is small, since then the additional jets from the
gluino decay will be soft. 
In the following we will only consider a scenario where the $m_{\go} -
m_{\sqL}$ mass difference is sufficiently large to allow a veto on
additional jets for gluino background reduction. 
We use a modification of the SPS1a scenario \cite{sps}, where 
the gluino mass is raised to 700 GeV. 

The most important backgrounds from Standard Model sources are 
$W^\pm W^\pm jj$, where $j$ is a light-flavor
jet, and semi-leptonic $t\bar{t}$, with the second lepton 
coming from the decay of a $b$ quark. Due to the large total
$t\bar{t}$ cross section, this can result in a sizable background. 

We compute numerical results for expected signal and background levels including some simple estimates for the detector response and resolution (see \cite{paper} for details), but do not perform a real experimental analysis. 
In our scenario, the chargino mainly decays into scalar taus, 
which subsequently decay into taus. 
To trace the charge explicitly, here we only consider 
the leptonic tau branching fraction in the decay chain
\begin{eqnarray}
\suL & \stackrel{65\%}{-\!\!\!-\!\!\!\longrightarrow} & d \, \cha^+_1
\stackrel{100\%}{-\!\!\!-\!\!\!\longrightarrow} d \, \tau^+ \, \nu_\tau \,
\neu_1 \stackrel{35\%}{-\!\!\!-\!\!\!\longrightarrow} d \, \ell^+ + \Eslash,
\qquad \ell = e, \, \mu,
\label{eq:dec}
\end{eqnarray}
and similarly for $\sdL$.
Both signal and top and gluino backgrounds were simulated with {\sc Pythia}
6.326 \cite{pythia}, while
the $WWjj$ background was generated with {\sc MadEvent} \cite{mad}. 
The cross sections for squark, gluino and top production were 
normalized with the $K$-factors for next-to-leading order QCD corrections
\cite{nlo}, while for the $W^\pm
W^\pm jj$ background only leading order results are available.

As a first step, the following preselection cuts are applied:
at least 100 GeV transverse missing energy, at least 2 jets with $p_{\rm T,j} > 100$  GeV, and two isolated leptons $\ell = e,\mu$ with $p_{\rm T,\ell}
  > 7$ GeV. At this level, most backgrounds are still larger
than the signal, see
Tab~\ref{tab:cuts}.
\begin{table}
\caption{Signal and background cross sections for progressive application of cuts.}
\label{tab:cuts}
\begin{tabular}{lrr|rrrrr}
Cross Sections & Signal & \multicolumn{5}{l}{Backgrounds} \\
$\sum_{q=u,d,s,c}\sigma$ (fb) & $\sqL\sqL$ & \bf Sum &
$t\bar{t}$ & $W^\pm W^\pm jj$ &  $\sqL\go$ & $\sqL\sqL^*$ & $\go\go$
\\
\hline
Total & 2100 & - & 8$\times 10^5$ & - & 7000 & 1350 & 3200
\\
\hline
Preselection    & 49.2 & 384.6 & 177.7 & - & 136.4 & 23.2 & 47.3
\\
b-veto          & 17.1 & 31.4 &  13.0 &  - & 10.3 & 7.1 & 1.0 
\\
$\Eslash > 150$ GeV &15.1 & 22.2 & 6.1 & - & 9.0 & 6.2 & 0.9 
\\
$p_{T,j_3}<50$ GeV& 7.8&5.9& 2.4 & N/A & 1.0 & 2.5 & 0.03
\\
$p_{T,j_1}>200$ GeV& \bf 7.0&\bf $<$4.9& 1.0 & $<$0.7 & 0.8 & 2.3 & 0.03
\\
\end{tabular}
\end{table}

Using a $b$ veto is effective against the gluino and $t\bar{t}$ backgrounds.
A high efficiency of $\epsilon = 90$\% reduces the background substantially, at the price of a also high mistagging $D=25$\% rate for the signal. The large SM backgrounds can  be further suppressed by a cut on the missing transverse energy $\Eslash$, with $\Eslash > 150$ GeV.
At this point, the gluino-related backgrounds dominate. 
They are reduced by a cut on hard additional jets.
By rejecting all events with $p_{\rm T,j3} > 50$~GeV, the ratio of the
signal to gluino background is markedly improved. 
Finally, increasing the transverse momentum cut on the first jet to $p_{\rm T,j1} >
200$ GeV, the top background is suppressed further, resulting in the cross section estimate in Tab.~\ref{tab:cuts}. The signal-to-background ratio is 1.4, 
sufficient to allow a meaningful
measurement. With an integrated luminosity of 100~fb$^{-1}$, the statistical
error on the same-sign squark cross section is 4.9\%.

\section{Squark decays at the ILC}
\label{sc:ilc}

In order to obtain from the measured rates at the LHC the total squark
production cross section,
the individual branching ratios (BRs) in the decay
chain eq.~\eqref{eq:dec} must be determined. Here we explore, how 
these could be extracted from measurements at ILC.

The chargino BRs can be determined from chargino pair production. Due to the large cross section for that process, all possible chargino decay channels can be easily separated from backgrounds, and the expected error on the BR is about 1\%.

In the given scenario the L-squarks are  slightly too heavy to 
be accessible at a 1 TeV linear collider. Here, we instead analyze the production of squarks for a hypothetical $e^+e^-$ collider with
a center-of-mass energy of about 1.5 TeV. 

The L-squarks can decay into the whole spectrum of charginos and neutralinos.
While in our scenario the light charginos and neutralinos decay into taus, the heavier states have large BRs into gauge bosons, and can be distinguished through these channels. See \cite{paper} for details.
The tau leptons in the final state can
be identified in their hadronic decay mode with roughly 80\% tagging
efficiency. 

For this work, Monte-Carlo samples for squark pair production in the different
squark decay channels have been generated at the parton level with the tools of
Ref.~\cite{slep}. Also the most relevant backgrounds have been
simulated, stemming from 
double and triple gauge boson production as well as $t\bar{t}$
production. It is assumed that an integrated luminosity of 500 fb$^{-1}$ is
spent for a polarization combination $P(e^+)$/$P(e^-)$ = +50\%/$-$80\%, which
enhances the production cross section both for $\suL$ and $\sdL$ production.
Here $\mp$ indicates left/right-handed polarization.
The BRs are obtained from measuring the
cross sections of all accessible decay modes of the squarks and identifying the
fraction of decays into one specific decay mode out of these. 

Since the squarks are produced in charge-conjugated pairs, it is a priori
difficult to distinguish up- and down-squarks in the final state. However,
assuming universality between the first two generations, a separation between
up- and down-type squarks can be obtained through charm tagging. According to
Ref.~\cite{ilctag}, a $c$-tagging efficiency of 40\% is achievable for a purity
of 90\%. By combining the different decay channels,
the following final state signatures are
identified as interesting:
$jj(n\tau)\Eslash$ with $n\in\{1,2,3,4\}$, 
$cc(n\tau)\Eslash$ with $n\in\{2,4\}$, 
$jj\tau\tau (Z/W) \Eslash$, $cc\tau\tau Z\Eslash$,
where $j$ indicates an untagged jet, $c$ a tagged charm jet, and $Z/W$ a
hadronically decaying gauge boson where the invariant mass of the two jets
combines to the given gauge boson mass. 
Since several squark decay channels can 
contribute to most of the final states above, 
one has to solve a linear equation system in order to derive
the individual contributions. 
With this procedure we estimate the precision for the BRs of the squarks into
charginos to be BR$(\suL \to d \cha^{\mbox{\tiny +}\!}_1) = (67.7 \pm 3.2)$\%
and BR$(\sdL \to u \cha^{\mbox{\tiny $-$}\!}_1) = (63.9 \pm 5.2)$\%.


\section{Combination and conclusions}
\label{sc:comb}

Based on the simulations for squark production at the LHC and
the ILC presented above, one can now derive an estimate for
the precision for the determination of the strong SUSY Yukawa
coupling $\hat{g}_{\rm s}$. The statistical uncertainty is combined with the most
important systematic error sources in Tab.~\ref{tab:res}.
\begin{table}
\caption{Combination of statistical and systematic errors for the
same-sign squark
cross section at the LHC and the derivation of the strong SUSY-Yukawa coupling.}
\label{tab:res}
\centering
\begin{tabular}{l@{\hspace{2cm}}rl}
 & $\sigma[\sqL\sqL]$ & $\hat{g}_{\rm s}/g_{\rm s}$ \\
\hline
LHC signal statistics & 4.9\% & 1.3\% \\
SUSY-QCD Yukawa coupling in $\sqL\go$ background & 2.4\% & 0.6\% \\
PDF uncertainty & 10\% & 2.4\% \\
NNLO corrections & 8\% & 2.0\% \\
Squark mass $\Delta m_{\sqL} = 9$ GeV & 6\% & 1.5\% \\
BR$[\sqL \to q' \, \cha^\pm_1]$ & 8.2\% & 2.0\% \\
\hline 
& 17.3\% & 4.1\% \\
\end{tabular}
\end{table}
We considered the following systematic error sources.
The remaining background from gluino production at the LHC introduces a
systematic error since it depends on $\hat{g}_{\rm s}$, which is estimated by varying $\hat{g}_{\rm s}$. The uncertainty from the proton parton distribution
functions (PDFs) is evaluated by comparing results for different CTEQ PDFs
\cite{cteq}. 
The uncertainty of the missing ${\cal O}(\alpha^2_{\rm s})$ radiative corrections are estimated by varying the renormalization scale of the ${\cal
  O}(\alpha_{\rm s})$ 
corrected cross section within a factor two. Furthermore the cross section depends on the values of the squark masses, which
according to Ref.~\cite{lhclc}
can be determined with an error better than $\Delta m_{\sqL} = 9$~GeV. Finally, the
expected error for the determination of the squark branching ratios at the
linear collider must be included.
Combining all error sources in quadrature, it is found that the SUSY-QCD Yukawa
coupling $\hat{g}_{\rm s}$ can be determined with an error 4.1\% in the given
scenario.

\end{document}